\documentclass[aps,prd,twocolumn,12pt,showpacs,showkeys,superscriptaddress,groupedaddress,reprint]{revtex4-1}
\pdfoutput=1
\usepackage{graphicx}  % needed for figures
\usepackage{dcolumn}   % needed for some tables
\usepackage{bm}        % for math
\usepackage{amssymb}   % for math
\usepackage{amsmath}   % for math
\usepackage{libertine}

\usepackage{color} 

\usepackage{multirow}
\usepackage{enumerate}
\usepackage{comment}
\usepackage{tabularx}

\newcommand{\drm}{{\rm d}}
\newcommand{\irm}{{\rm i}}
\newcommand{\erm}{{\rm e}}
\newcommand{\beq}{\begin{equation}}
\newcommand{\eeq}{\end{equation}}
\newcommand{\bdm}{\begin{displaymath}}
\newcommand{\edm}{\end{displaymath}}

\begin{document}

\title{Searching for cosmological gravitational-wave backgrounds with third-generation detectors in the presence of an astrophysical foreground}

\author{Ashish Sharma}
\email{ashish.sharma@gssi.it}
\affiliation{Gran Sasso Science Institute (GSSI), I-67100 L'Aquila, Italy}
\affiliation{INFN, Laboratori Nazionali del Gran Sasso, I-67100 Assergi, Italy}
\author{Jan Harms}
\email{jan.harms@gssi.it}
\affiliation{Gran Sasso Science Institute (GSSI), I-67100 L'Aquila, Italy}
\affiliation{INFN, Laboratori Nazionali del Gran Sasso, I-67100 Assergi, Italy}

\date{\today}

\begin{abstract}
The stochastic cosmological gravitational-wave background (CGWB) provides a direct window to study early universe phenomena and fundamental physics.  With the proposed third-generation ground-based gravitational wave detectors, Einstein Telescope (ET) and Cosmic Explorer (CE), we might be able to detect evidence of a CGWB. However, to dig out these prime signals would be a difficult quest as the dominance of the astrophysical foreground from compact-binary coalescence (CBC) will mask this CGWB. In this paper, we study a subtraction-noise projection method, making it possible to reduce the residuals left after subtraction of the astrophysical foreground of CBCs, greatly improving our chances to detect a cosmological background. We carried out our analysis based on simulations of ET and CE and using posterior sampling for the parameter estimation of binary black-hole mergers. We demonstrate the sensitivity improvement of stochastic gravitational-wave searches and conclude that the ultimate sensitivity of these searches will not be limited by residuals left when subtracting the estimated BBH foreground, but by the fraction of the astrophysical foreground that cannot be detected even with third-generation instruments, or possibly by other signals not included in our analysis. We also resolve previous misconceptions of residual noise in the context of Gaussian parameter estimation.
\end{abstract}

\maketitle

\section{Introduction}  

The accomplishment of detecting gravitational waves (GWs) from the mergers of compact binaries with neutron stars and black holes opened a new window to study astrophysical and cosmological phenomena of the Universe. The continuous improvement in the sensitivity and multi-detection of signals due to coalescence of a binary neutron star (BNS) and various binary black-hole (BBH) mergers during the first two observation runs of Advanced LIGO \cite{aLIGO} and Advanced Virgo \cite{Virgo} marks the beginning of a cosmic catalog of sources so far reaching out to distances of about 3\,Gpc and only capturing a small fraction of all compact binaries in this volume \cite{AbbottMerger2019}.  

A major objective of modern cosmology is to detect early-universe GW signals, which are crucial to test current cosmological models and to further our understanding of the evolution of the Universe \cite{AbEA2009,AdEA2018}.  The cosmic GW background (CGWB) is predicted to arise from fundamental processes in the early universe \cite{Caprini2018,Nelson2018}. Among these are quantum vacuum fluctuations amplified by inflation \cite{Turner1997, Easther2007,  Guzzetti2016},  phase transitions \cite{Marc1994, Tina2008, Caprini2009,  Caprini2016}, and also cosmic strings are prominent for CGWB searches with ET and CE \cite{Tanmay1985, Damour2005, Xavier2007, Olmez2010}. On the theoretical side, there is huge advancement to understand the concept and generation of these cosmological signals and on the observational and experimental side, to detect these signals with GW detectors is also in advancement and provides us with the capability to detect these signals in the future. 

However, the detection of a CGWB is extremely challenging. Mission concepts that would make the detection of a primordial stochastic background probable, the space-borne detectors Big-Bang Observer (BBO) \cite{BBO} and DECIGO \cite{DECIGO}, still require substantial advances in laser-interferometer technology, and it is unknown when or if these experiments will become operational. Two ground-based, third-generation GW detectors have been proposed, Einstein Telescope (ET) \cite{ET} and Cosmic Explorer (CE) \cite{CE1}, which are expected to be operational by 2035 and potentially with the capacity to detect a CGWB. Primordial stochastic signals are predicted to lie well below instrument noise of all conceived future GW detectors. The stochastic searches with GW detectors follow the cross-correlation method betting on the assumption that fields of stochastic GWs produce correlations between detectors, while the instrument noises do not, or in a well-understood way with options to mitigate correlated noise, e.g., by Schumann resonances \cite{TCS2013,CoEA2016}. Optimal cross-correlation filters can be employed to obtain the maximum signal-to-noise ratio (SNR) integrated over a band of frequencies and thus maximize chances to detect a CGWB \cite{Chr1992,Bruce1999}. It is also possible to estimate parameters of stochastic backgrounds such as spectral slopes and possible anisotropies of the GW field \cite{Bal2006,Romano2019}. 

With the proposed third-generation GW detectors ET and CE, we will step into a new era of GW physics, and we will overcome the scarcity of GW sources, such that we will be able to detect binary signals up to high redshifts $z \geq10$. Analyses of data from the first and second observation runs of LIGO and VIRGO constrain the local BBH merger rate to about 10 -- 100\,Gpc$^{-3}$y$^{-1}$ \cite{AbbottMerger2019}. The BBH merger rate as a function of redshift is estimated from the star-formation rate \cite{VaEA2015}, distribution of time-delays between formation and merger \cite{BKB2002}, and by normalizing to the local merger rate \cite{Vitale2019}. It predicts about $10^{5}-10^{6}$ BBH mergers per year and a large fraction of them detectable with ET or CE. Since the correlation between detectors is predicted to be dominated by the astrophysical foreground of compact-binary coalescences (CBCs), detection of a cosmological background is strongly impeded and mitigation of the foreground is required.  

As a first step, the foreground can be reduced by subtracting the estimated waveforms of all detected signals. Previous work has shown that a combination of unresolved sources, i.e., signals lying below the detection threshold, and residuals left in the data after subtraction can still limit the sensitivity of CGWB searches with future GW detector networks \cite{Tania2017,SRS2020}. Both publications neglected the possibility to reduce subtraction residuals as proposed in earlier work for space-borne detectors \cite{Curt2006,Jan2008}. Furthermore, a full-Bayesian analysis of primordial and astrophysical signals is expected to lower the impact of sub-threshold signals \cite{SmTh2018}. In this paper, we deal with the problem of reducing the subtraction residuals of the astrophysical foreground in third-generation detectors ET and CE, and for this goal we test the subtraction-noise projection method for BBHs \cite{Curt2006,Jan2008}. As was pointed out in a recent publication \cite{SRS2020}, the foreground of binary neutron stars (BNS) is expected to be more challenging to reduce, but we have no means yet to simulate this problem beyond what has already been done in previous work, as it requires an effective posterior sampler for parameter estimation in ET, which is a major challenge due to the length of BNS waveforms in ET (around a day), and it needs to account for the rotation of Earth during observation time.

The paper is organized as follows. Details of the simulation of the detector network and its astrophysical foreground are presented in section \ref{sec:SimOV}. Section \ref{sec:geometry} reviews the geometrical interpretation of matched filtering, and provides an estimate of residual noise from an astrophysical foreground. In section \ref{sec:PSN}, we explain the projection method that reduces residuals of the astrophysical foreground. The cross-correlation measurement between CE and ET is detailed in section \ref{sec:SBD}. Results of the foreground-mitigation procedure are discussed in section \ref{sec:results}.

\section{Simulation Overview} 
\label{sec:SimOV}

\begin{figure}[htp]
	\centering
	\includegraphics[width=1\linewidth]{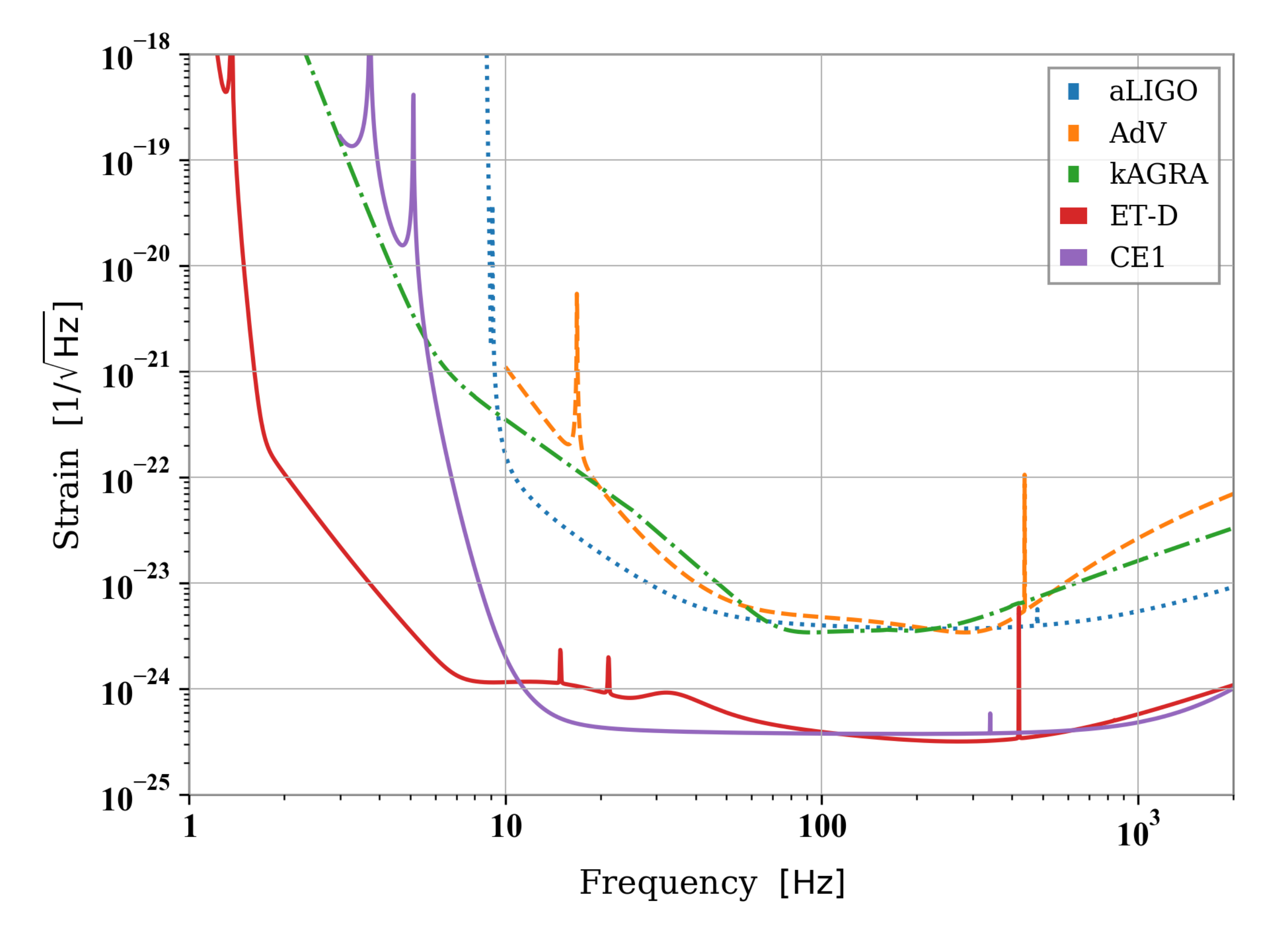}
	\caption{Design sensitivities of current and future GW detectors.}
	\label{fig:SenCurve}
\end{figure}

The second-generation detectors Advanced LIGO and Advanced Virgo, after gradual updates, already observed several tens of binary-merger signals including the candidates of the last observing run \cite{AbbottMerger2019}. There is still a huge spectrum of GW physics unexplored both in astrophysics and cosmology. In future, the main focus will be to exploit this vast spectrum. To progress in this direction, we need next-generation GW detectors with much better sensitivity than current GW detector. Two ground-based detectors have been proposed so far: the European Einstein Telescope and the US Cosmic Explorer \cite{ET,CE}. Individually and as a detector network together with developed versions of current-generation detectors (including KAGRA \cite{AkEA2018} and LIGO India \cite{Sou2016}), these third-generation detectors have a rich science case covering topics in fundamental physics, cosmology, astrophysics, and nuclear physics \cite{3Gscience,ETscience}. Their projected sensitivities are shown in figure \ref{fig:SenCurve}.

\textbf{ET}:  ET is a proposed European third-generation, underground GW observatory in the shape of an equilateral triangle with 10\,km side length. ET will provide an improvement in sensitivity by a factor of 10 with respect to current GW detectors, extending the observation band down to about 3\,Hz \cite{HiEA2011}. The Einstein Telescope will be placed underground to reduce the environmental noise coming from seismic and atmospheric fields. The infrastructure will host three interferometer pairs, each pair consisting of a low-frequency and a high-frequency interferometer forming a so-called xylophone configuration \cite{HiEA2009}.

\textbf{CE}: CE is a proposed US third-generation, surface GW observatory with the traditional L-shape and arm length of 40\,km. Its design also foresees a sensitivity improvement by about a factor 10 compared to current GW detectors. The sensitivity model employed in this study corresponds to the first phase of CE development (CE1 in \cite{CE1}). Its ultimate sensitivity target is about a factor 2 better than this.

The basis of our simulation is the calculation of a 1.3-year-long stretch of GW data for ET (three individual data streams) and CE. The subtraction of best-fit waveforms is carried out in time domain, while the residual-noise projection is easiest to perform in the domain of the waveform model (frequency domain in our case) as explained in section \ref{sec:PSN}. The projection requires Fisher matrices, which in turn require the derivatives of waveforms with respect to their parameters. We carry out the differentiation numerically so that in future, we can use this simulation also to study systematics related to waveform modeling without requiring analytic waveforms. Cross-spectral densities (CSDs) of time series between all four detectors are calculated after each of the following steps,
\begin{enumerate}
\item creation of time series only with instrument noise,
\item injection of GW signals in all four detectors (3 ET + 1 CE),
\item subtraction of best-fit waveforms,
\item residual-noise projection,
\end{enumerate}
to demonstrate the impact of each step on the CSD. Finally, optimal filters are applied for an evaluation of the ultimate sensitivity of the network to a CGWB. 

Whenever possible, our analysis uses functions of the Python parameter-estimation software package \emph{Bilby} \cite{AsEA2019}. The calculation of noise time series is done by built-in functions of Bilby using instrument noise models of ET and CE in the form of spectral densities. Also injection of GW signals in the data, and posterior sampling are done with Bilby. Subtraction of the best-fit waveforms is done using the injection algorithm with a change of the sign of the waveform. The projection of residual noise is mostly based on original code. The optimal filters used in the final step for the detection of a CGWB depend on the overlap-reduction function between detectors \cite{Chr1992,Bruce1999}, which can be calculated straight-forwardly using antenna patterns provided by Bilby.

The astrophysical foreground is formed by the mergers of compact binaries with black holes and neutron stars. The lowest-mass members of these binaries, especially the binary neutron stars, take a special role (from today's perspective) since it is still prohibitively expensive to simulate parameter estimation accurately for ET given that these signals can last for more than a day in the ET observation band and generally require high sampling frequencies to study the merger physics. A multi-band analysis of individual signals might provide a solution \cite{AdEA2016,VVM2017}, but there is no parameter-estimation package yet based on posterior sampling and implementing all required effects such as the impact of Earth's rotation on the GW signal. For this reason, we chose to focus on BBH mergers in this paper. It allows us to use state-of-the-art parameter estimation software for posterior sampling.

\begin{figure}[htp]
	\centering
	\includegraphics[width=0.7\linewidth, height=0.7\linewidth]{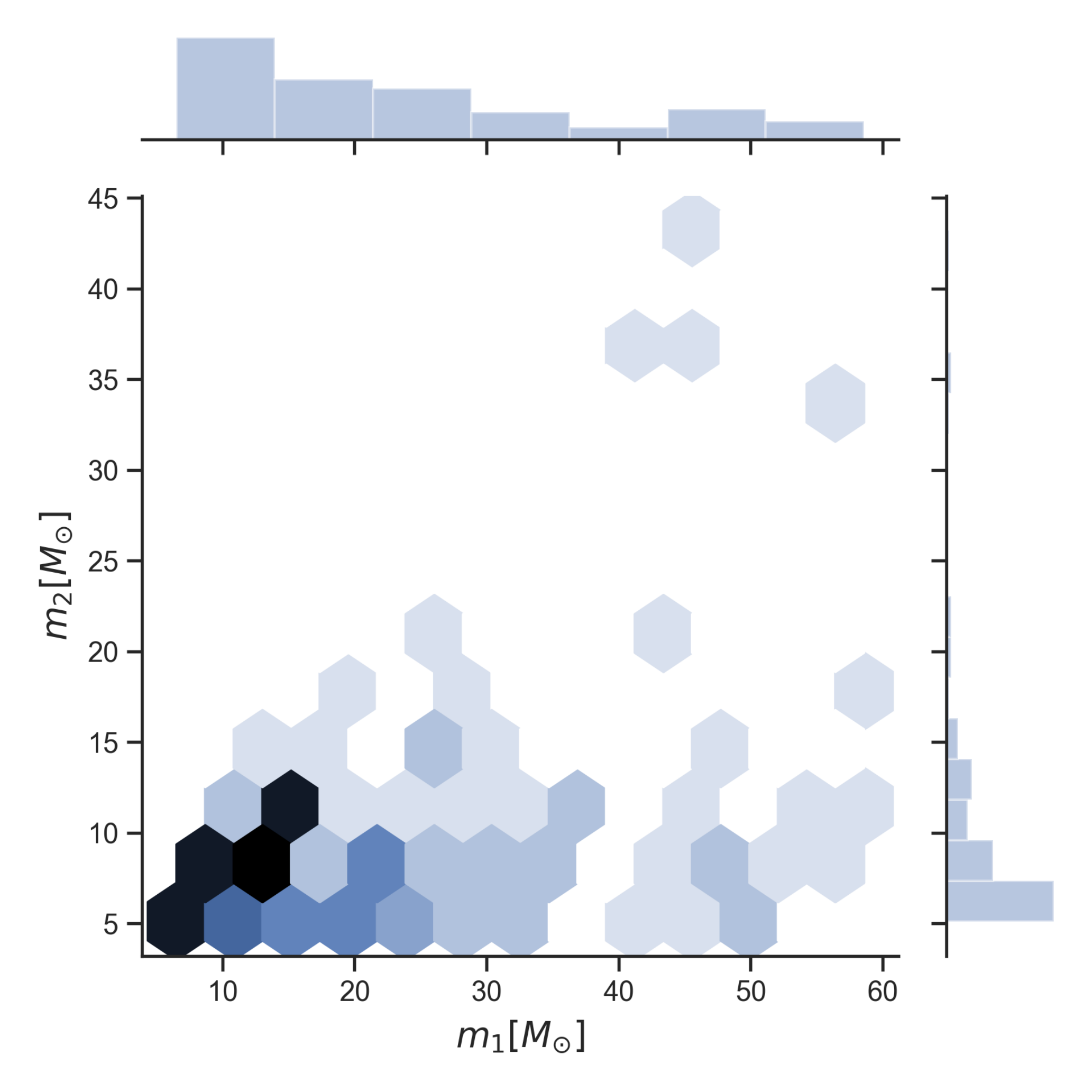}
	\caption{Mass range for individual BBH mass used in our study.}
	\label{fig:masses12}
\end{figure}

\begin{figure}[htp]
	\centering
	\includegraphics[width=0.7\linewidth, height=0.7\linewidth]{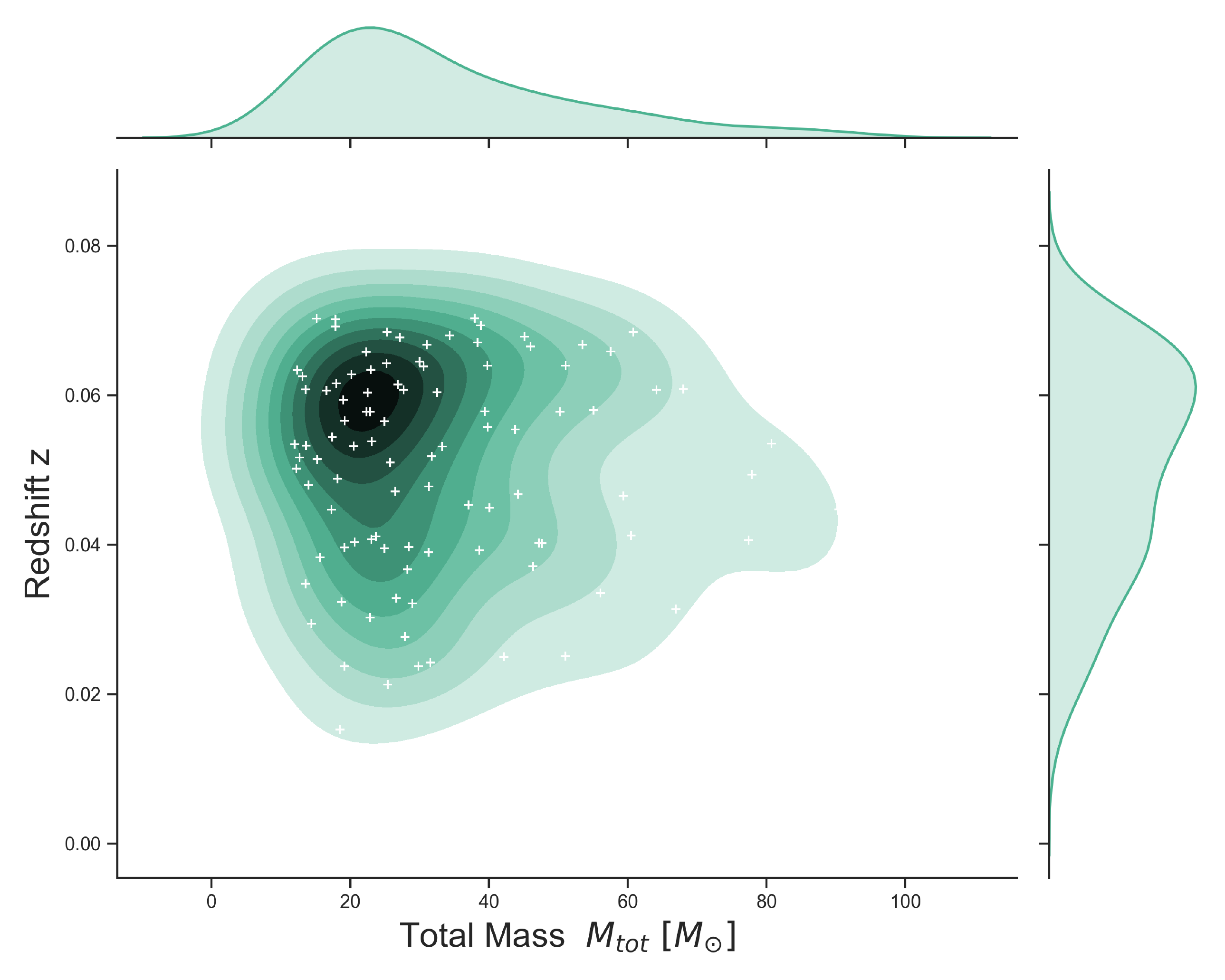}
	\caption{Distribution of total mass $M$ and redshift $z$ of the 100 BBH signals used in our analysis. High-SNR signals are chosen for clearest possible demonstration of the projection method to mitigate the astrophysical foreground.}
	\label{fig:TotMassRedS}
\end{figure}

However, even when focusing on BBH mergers, providing parameter estimations of $10^{5}-10^{6}$ signals by posterior sampling, which is an important new ingredient in this work compared to previous studies of the projection method, is computationally prohibitively expensive. As a way forward, we adopted the following scheme. Only for 100 BBH signals, posterior sampling is performed. The complete stretch of data is divided into 10000 segments of length 4096\,s. In each of the segments, all 100 waveforms are injected with random time shifts so that the merger occurs in the respective time segment. In this way, phase relations between all signals are randomized and the CSD between detectors has the properties of a stochastic foreground. Its overall amplitude is \emph{stronger} than it would be in a more realistic simulation since we deliberately chose highest-SNR members of the cosmological distribution to have the clearest demonstration of the effect of residual-noise projection.  We employ a redshift independent power-law distribution for both intrinsic masses with a power law index $\alpha = -1.6$ \cite{Abbott2019} constraining the individual masses to lie within the range $5 \mathcal{M}_{\odot} \leq m_{2}  < m_{1}  \leq 60 \mathcal{M}_{\odot}$. This leads to the sample of 100 BBH masses shown in figure \ref{fig:masses12}. 
The sampled redshifts and total masses of the signals used in this paper are shown in figure \ref{fig:TotMassRedS} together with smoothed distributions derived from these samples (which explains why there is no low-mass bound of the mass distribution).

\section{Matched filtering and the residual of an astrophysical foreground} 
\label{sec:geometry}

The probability of two BBH signals to overlap in time in ET data is relatively high, but depends on details of the mass distribution \cite{Tania2017}. Lower-mass signals last longer (up to a few minutes) and if present in greater number would lead to more frequent overlap. However, it is unlikely that the number of detectable BBH signals with 3G detectors will be impacted significantly by the presence of other signals, which is in contrast to the situation described for future space-borne GW detectors where the foreground acts as excess noise \cite{LISA,Curt2006}. It is therefore enough to consider the impact of the astrophysical foreground on the correlation measurements between detectors, which is addressed by the subtraction-projection method discussed in the following using results from Cutler and Harms \cite{Curt2006}.

The basis of the subtraction-projection method is the expansion of parameter errors or likelihood functions with respect to the inverse of the SNR of signals, which means that this approach works better for high-SNR signals. An important quantity is the Fisher matrix, whose components take the form
\begin{equation} 
\begin{split}
\Gamma_{\alpha \beta} &= <\partial_{\alpha} h(\vec\lambda) | \partial_{\beta}h(\vec\lambda) > \\
&= 4 \int_{0}^{\infty} df \frac{\Re(\partial_{\alpha} h(f) \partial_{\beta} h^{*}(f))}{S^{n}(f)},
\end{split}
\label{eq:fisher}
\end{equation}
where $\vec\lambda$ is the vector of model parameters. The scalar product requires an estimate of the instrument-noise spectral density $S^{n}(f)$. We expressed the Fisher matrix as a scalar product $\langle\cdot|\cdot\rangle$ between derivatives of the waveform model with respect to model parameters $\partial_\alpha=\partial/\partial\lambda^\alpha$. The Fisher matrix can be interpreted as a metric on the curved template manifold defined by the waveform model $h(\vec\lambda)$. The template manifold is a sub-manifold of the sampling space whose points describe realizations of detector data including instrument noise and signals not described by $h(\vec\lambda)$. 

If the best-fit waveform $\hat h$ maximizes the likelihood (standard parameter estimation maximizes the posterior that includes priors), i.e., if it minimizes $\langle s-\hat h|s-\hat h \rangle$, then for signals with sufficiently high SNR, $\hat h$ fulfills the following equation
\beq
\langle s-\hat h|\partial_\alpha\hat h\rangle=\langle n+(h-\hat h)|\partial_\alpha\hat h\rangle=0
\label{eq:bestfit}
\eeq 
for all derivatives $\partial_\alpha$, with $n$ being the instrument noise and $h$ the GW signal contributing to the data $s$. The vanishing of the first scalar product in this equation means that the line in sampling space connecting the point $s$ with the best-fit waveform $\hat h$ is perpendicular to the template manifold, i.e., the best-fit waveform is obtained by determining the template on the manifold with minimal distance to $s$. The vanishing of the second scalar product means that the residual noise $\delta h=\hat h-h$ is equal to the component of the instrument noise tangent to the manifold at the point of the best-fit $\hat h$. 

The scalar product can also be used to define the SNR of a signal:
\beq
{\rm SNR}=\sqrt{\langle  h|h\rangle}
\label{eq:SNR}
\eeq
The leading order term of a SNR$^{-1}$ expansion of the covariances of parameter-estimation errors $\delta\lambda^\alpha$ is given by 
\beq
\overline{\delta\lambda^\alpha\delta\lambda^\beta}=\Gamma^{\alpha\beta},
\label{eq:avPEerrors}
\eeq
where $\Gamma^{\alpha \beta}$ are the components of the inverse of the Fisher matrix. This relation is sometimes used to define an approximate Gaussian distribution of the likelihood function $\exp(-\Gamma_{\alpha\beta}\delta\lambda^\alpha\delta\lambda^\beta/2)$, and parameter-estimation errors can be drawn from this distribution to substitute a computationally costly posterior sampling \cite{Jan2008,SRS2020}.

It is also possible to express the leading-order, parameter-estimation errors $\delta \lambda^\alpha$ in terms of a specific instrumental-noise realization $n$ as:
\begin{equation} 
\delta \lambda^{\alpha} \equiv\hat\lambda^\alpha -\lambda^\alpha \approx \Gamma^{\alpha \beta} <n|\partial_{\beta} \hat{h}>.
\label{eq:PEerrors}
\end{equation}
Here, $\hat\lambda^\alpha$ are the parameter estimates determining the best-fit waveform $\hat h$. By using equation (\ref{eq:avPEerrors}), we can calculate the norm-squared of the average subtraction residuals
\begin{equation}
\overline{<\delta h|\delta h>} = <\partial_{\alpha} \hat{h}| \partial_{\beta} \hat{h}> \overline{\delta \lambda^{\alpha} \delta \lambda^{\beta}} = \Gamma_{\alpha \beta} \Gamma^{\alpha \beta} =N_p ,
\label{eq:13}
\end{equation}
where $N_p$ is the total number of parameters going into the waveform model $h$. Together with equation (\ref{eq:SNR}) it tells us that in average, the amplitude of a signal after subtraction of its best-fit waveform is reduced by $\delta h/h\sim N_p^{1/2}/\rm SNR$, which also means that the residual is independent of the SNR of the signal (again, in the approximation of large SNR).

With the future GW detectors ET and CE we will be able to detect almost all the BBHs emitting within their observation bands, and the entire astrophysical foreground coming from $N_S$ sources is
\begin{equation} 
H(t)=\sum_{k = 1}^{N_{\rm S}} h^k(t).
\label{eq:sumh}
\end{equation}
With each BBH signal $h^k(t)$ being described by $N_p$ parameters, the parameter space of the complete astrophysical foreground has dimension $N_p\times N_S$. Therefore the norm-squared of the residual of this foreground is
\begin{equation} 
\overline{<\delta H|\delta H>} = N_p \times N_S.
\label{eq:resH}
\end{equation}
It is easy to show that in average
\begin{equation}
\delta H / H  \approx \delta h/h,
\end{equation}
which means that the fractional reduction of the amplitude of a single BBH is about the same of the entire astrophysical foreground assuming that (almost) all signals can be detected with sufficiently high SNR.

\section{Projecting out the residual noise}
\label{sec:PSN}
The results of section \ref{sec:geometry} form the basis of the residual-noise projection, which we discuss in the following. As shown in equation (\ref{eq:bestfit}), the residual noise is tangent to the waveform manifold. The strategy of the projection method is to apply a projection operator to the residual data $r=s-\hat h$ removing all of its components lying in the manifold's tangent space at the best-fit waveform. This projection needs to be done for all the signals in the data. The projection operator can be written
\begin{equation} 
P \equiv 1 - \Gamma^{\alpha \beta} |\partial_{\alpha} \hat h><\partial_{\beta} \hat h|.
\label{eq:projection}
\end{equation}
When applying the projection to the residual data of a detector $i$, one obtains
\begin{equation} 
P[r_i](t) = r_i(t) -\Gamma_i^{\alpha \beta} \langle \partial_{\beta} \hat h_i| r_i\rangle \partial_{\alpha} \hat h_i(t),
\label{eq:projres}
\end{equation}
which we wrote here for the time domain, but it can also be applied in Fourier domain. The residual $P[r_i]$ after projection corresponds to the instrument noise perpendicular to the template manifold plus a potential component of the true signal $h_i$ that does not lie in the tangent space of the best-fit $\hat h_i$. This residual of the signal is non-vanishing only for curved manifolds and is suppressed by SNR$^{-2}$ relative to the original signal. Comparing with equation (\ref{eq:bestfit}), it seems that the projection operator should not have any effect on the residual data $r_i=s_i-\hat h_i$ since the vector $|r_i\rangle$ is normal to the tangent space of the template manifold at the best-fit. However, this is not necessarily correct for various reasons. 

First, the maximum-likelihood parameter estimates $\hat \lambda^\alpha$ are obtained using data from all detectors in the network. These parameter values determine the best-fit waveforms $\hat h_i=h_i(\hat \lambda^\alpha)$ of each detector $i$ in the network. These waveforms however are \emph{not} the results of a normal projection of data vectors $|s_i\rangle$ onto the respective template manifolds. This would only be the case if maximum-likelihood estimates $\hat \lambda^\alpha_i$ are calculated for each detector separately. This means that subtracting $h_i(\hat \lambda^\alpha)$ from the data of all detectors leaves residuals in the tangent spaces, which can be projected out. This also means that one needs to distinguish between the Fisher matrices $\Gamma_{i,\alpha\beta}(\hat\lambda_i^\mu)$ and $\Gamma_{i,\alpha\beta}(\hat\lambda^\mu)$, where the latter is obtained using the parameter estimates from a coherent network analysis. 

Let us consider the case where the maximum-likelihood estimations are done for each detector separately producing different best-fit parameters $\hat \lambda^\alpha_i$ for each detector $i$. Then, subtracting $\hat h_i=h_i(\hat \lambda^\alpha_i)$ for all signals in the data reduces the astrophysical foreground by 1/SNR$^2$ instead of 1/SNR. One might wonder where the subtraction residuals at order 1/SNR are, since clearly the misfit $\delta h_i$ is still only suppressed by $1/$SNR compared to the true signal $h_i(\lambda_0^\alpha)$. Here, the important point is that when subtracting a signal, the residual $\delta h_i$ is already exactly canceled by the component $n_\parallel$ of the instrument noise that lies in the tangent space, which can be understood from equation (\ref{eq:bestfit}) when using $n=n_\perp+n_\parallel$ and therefore $r=n_\perp$ plus residual noise from the astrophysical foreground suppressed by $1/$SNR$^2$ and higher.

Another reason why best-fit residuals can be in tangent spaces of a template manifold, even if the best-fits are calculated for each detector individually, is that they are typically not the result of a likelihood maximization, but of a maximization of the posterior distribution, which depends on priors. In this case, the residual $\delta h$ does not fulfill equation (\ref{eq:bestfit}), and residuals in tangent spaces remain to be projected out.

\begin{figure}[htp]
	\centering
	\includegraphics[width=1\columnwidth]{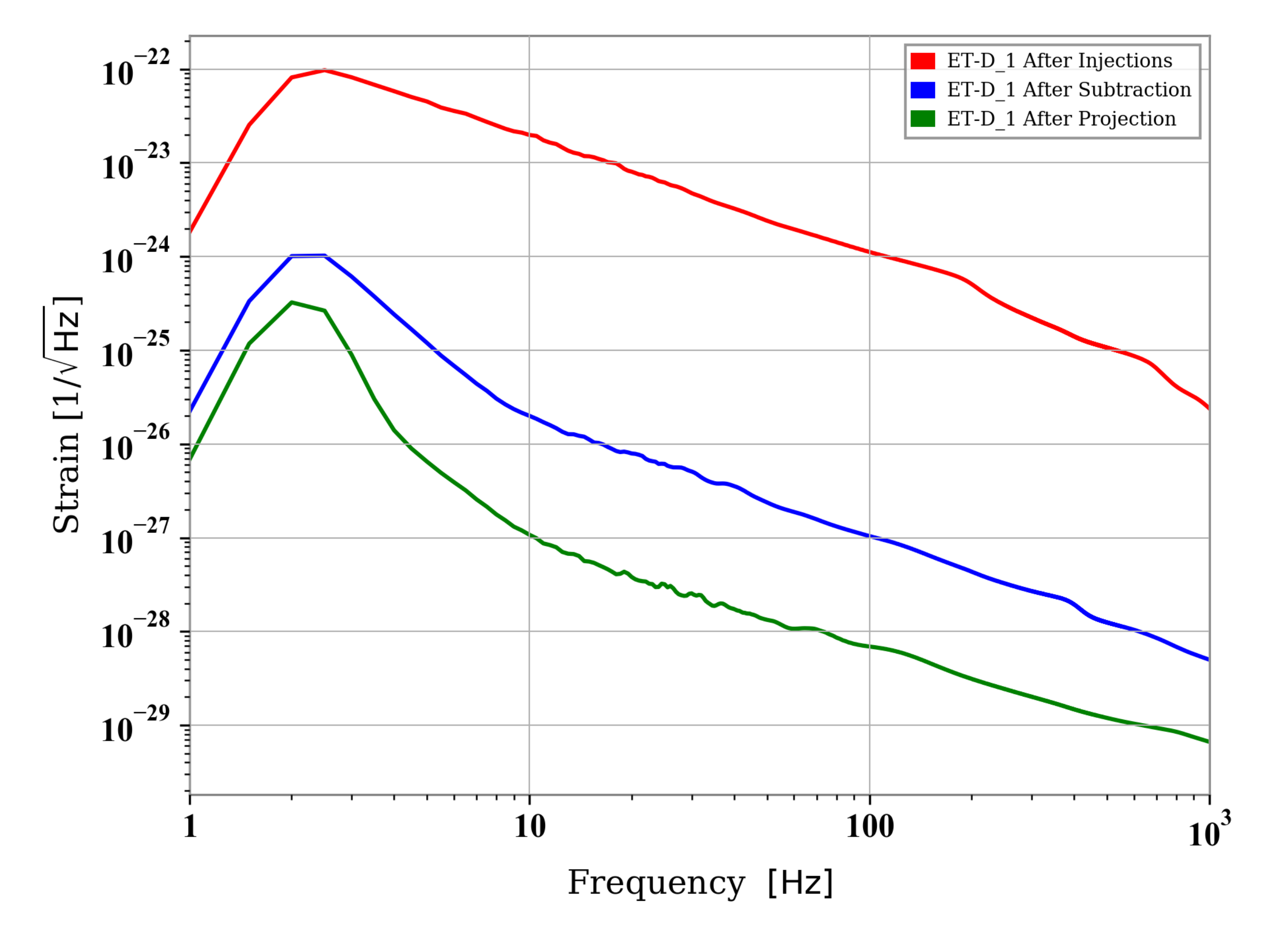}
	\caption{Simulated astrophysical foreground, subtraction residuals, and residual spectrum after projection for ET without instrument noise (except for the max-posterior parameter estimation).}
	\label{fig:resspec}
\end{figure}
Finally, technical choices of a simulation can lead to additional residuals in tangent spaces. Often, parameter estimation by posterior sampling is computationally too expensive for studies with a large population of signals. In this case, past work made use of equation (\ref{eq:avPEerrors}) to define a Gaussian error distribution, from which parameter errors are drawn and added to the true signal parameters to obtain the maximum-likelihood parameters \cite{Jan2008,SRS2020}. The issue here is that the parameter errors are not consistent with a specific realization of the instrument noise. The best-fit waveforms obtained in this way would not maximize the likelihood, and this leads to excess residual noise in tangent spaces, which is projected out \cite{Jan2008}. This artifact can be avoided by using equation (\ref{eq:PEerrors}) to obtain parameter errors, which is still under the assumption of a Gaussian likelihood, but at least consistent with a specific noise realization.

As a first demonstration, we show the root power-spectral density of the astrophysical foreground averaged over 1.3 years, its subtraction residual, and the spectrum after projection in figure \ref{fig:resspec} without instrument noise for an ET detector. For this plot, time series were simulated without instrument noise just to demonstrate the full potential of the projection method. The posterior sampling was of course done including instrument noise, and included data from CE and the full ET triangle. The simulated astrophysical foreground is artificially enhanced to make sure that all signals have sufficiently high SNR to be able to neglect residuals at order 1/SNR$^2$. Furthermore, one needs to consider the possibility that some low-SNR signals are not detected by ET, which gives rise to additional contributions to residual noise that we do not consider in this study (see instead \cite{Tania2017,SRS2020}). It is interesting to observe that the spectra change their shape after applying the subtraction and projection, for which we cannot provide an explanation since our equations only  predict residuals integrated over all frequencies.

\section{Stochastic background and Detection}
\label{sec:SBD}

The fractional energy-density spectrum of an isotropic stochastic background is defined as
\begin{equation}
\Omega_{GW} \left(f\right) ~=~ \frac{1}{\rho_{\rm c}} \cdot \frac{{\rm d} \rho_{\rm GW}}{{\rm d}\ln f},
\label{eq:Omega}
\end{equation}
where $\rho_{\rm c} ~=~ 3H_{0}^{2}c^{2}/\left(8\pi G\right)$ is the critical energy density required for a flat universe, $H_{0}$ is the Hubble constant ($H_{0} = 67.9\,\rm km\,s^{-1} Mpc^{-1}$  \cite{Planck2015}) and $\drm\rho_{\rm GW}$ is the energy density of GWs contained in the frequency band $f$ to $f+\drm f$ \cite{Bruce1999}.  The current limit on the gravitational-wave energy density spectrum is $\Omega_{\rm GW}  < 4.8 \times 10^{-8}$ with $95\%$ confidence, in the band 20--100\,Hz  \cite{AbbottCosmo2019}. In this work, we simulate searches optimized for an unpolarized, isotropic, stationary and Gaussian stochastic background. In reality, stochastic signals do not necessarily have these properties \cite{Bruce1999} except for stationarity, which is simply a consequence of short observation time compared to time scales characteristic for the evolution of GW distributions.

\subsection{Cross-Correlation between Detectors}
\label{subsec:CCP}
Cross-correlating the output of two or more GW detectors is the optimal strategy to detect a Gaussian, stationary stochastic GW background \cite{Chr1992,Bruce1999}. Since we prefer to work in frequency domain, the cross-correlation is expressed as cross power-spectral density (CPSD) $C_{ij}(f)$ between two detectors $i,\,j$. We briefly review the steps to calculate the contribution of an isotropic, stochastic GW background to $C_{ij}(f)$ and how to calculate the statistical error due to instrument noise.

A stochastic GW background can be described as a plane-wave expansion of a metric perturbation 
\begin{equation}
\begin{split}
h_{\mu\nu}&(\vec x,t) = \\
&\sum\limits_{A=+,\times}\int\limits_{s^{2}} \drm\hat{\Omega}\int\limits_{-\infty}^\infty\drm f\,h_{A}(f,\hat{\Omega})\erm^{\irm 2\pi f(t-\hat{\Omega}\cdot\vec{x}/c )} e^{A}_{\mu\nu}(\hat{\Omega}).
\end{split}
\label{eq:GWFourier}
\end{equation}
Here, $\hat\Omega$ is a unit vector pointing along the propagation direction of a GW, $c$ is the speed of light, $A$ is the wave polarization, $e^{A}_{\mu\nu}(\hat{\Omega})$ the polarization tensor, and $h_{A}(f,\hat{\Omega})$ the amplitudes of the plane waves.

The CPSD can now be calculated between two detectors at locations $\vec x_i,\,\vec x_j$ and antenna patterns
\begin{equation}
F^{A}_{i}(\hat{\Omega}) = e^{A}_{\mu\nu} d^{\mu\nu}_{i} = e^{A}_{\mu\nu} \frac{1}{2} (\hat{X}^{\mu}_{i}\hat{X}^{\nu}_{i} - \hat{Y}^{\mu}_{i}\hat{Y}^{\nu}_{i}),
\label{eq:antenna}
\end{equation}
where $X_i^\mu,\,Y_i^\mu$ are components of the unit vectors along the two arms of detector $i$, which define the components of the response tensor $d^{\mu\nu}_i$ of the detector. Even though the notation $X,\,Y$ suggests that arms are perpendicular to each other, this does not need to be the case (as for ET). Assuming that plane-wave contributions to the metric in equation (\ref{eq:GWFourier}) at different frequencies, from different directions, and different polarization are uncorrelated, the CPSD can be calculated in a straight-forward manner. The dependence of the CPSD on detector positions and orientations is summarized in the so-called overlap-reduction function (ORF) \cite{Chr1992,Bruce1999,Nishizawa2009}
\begin{equation}
\gamma_{ij}(f) = \frac{5}{8\pi} \sum_{A} \int\limits_{S^{2}} \drm\hat{\Omega} \erm^{\irm 2\pi f \hat{\Omega} \cdot \Delta\vec{x}_{ij}/c} F^{A}_{i}	(\hat{\Omega}) F^{A}_{j}(\hat{\Omega}).
\label{eq:ORF}
\end{equation}
Since the stochastic background is assumed to be homogeneous, $\gamma_{ij}$ only depends on the relative position vector $\Delta\vec x_{ij}=\vec x_j-\vec x_i$ between the two detectors. The numerical constant $5/(8\pi)$ is chosen such that $\gamma_{ij}=1$ for two detectors that are collocated, co-aligned and both having perpendicular arms. Even for GW detectors with non-perpendicular arms like ET, it is convenient to adopt the same normalization of the ORF. 

\begin{figure}[htp]
	\centering
	\includegraphics[width=1\linewidth]{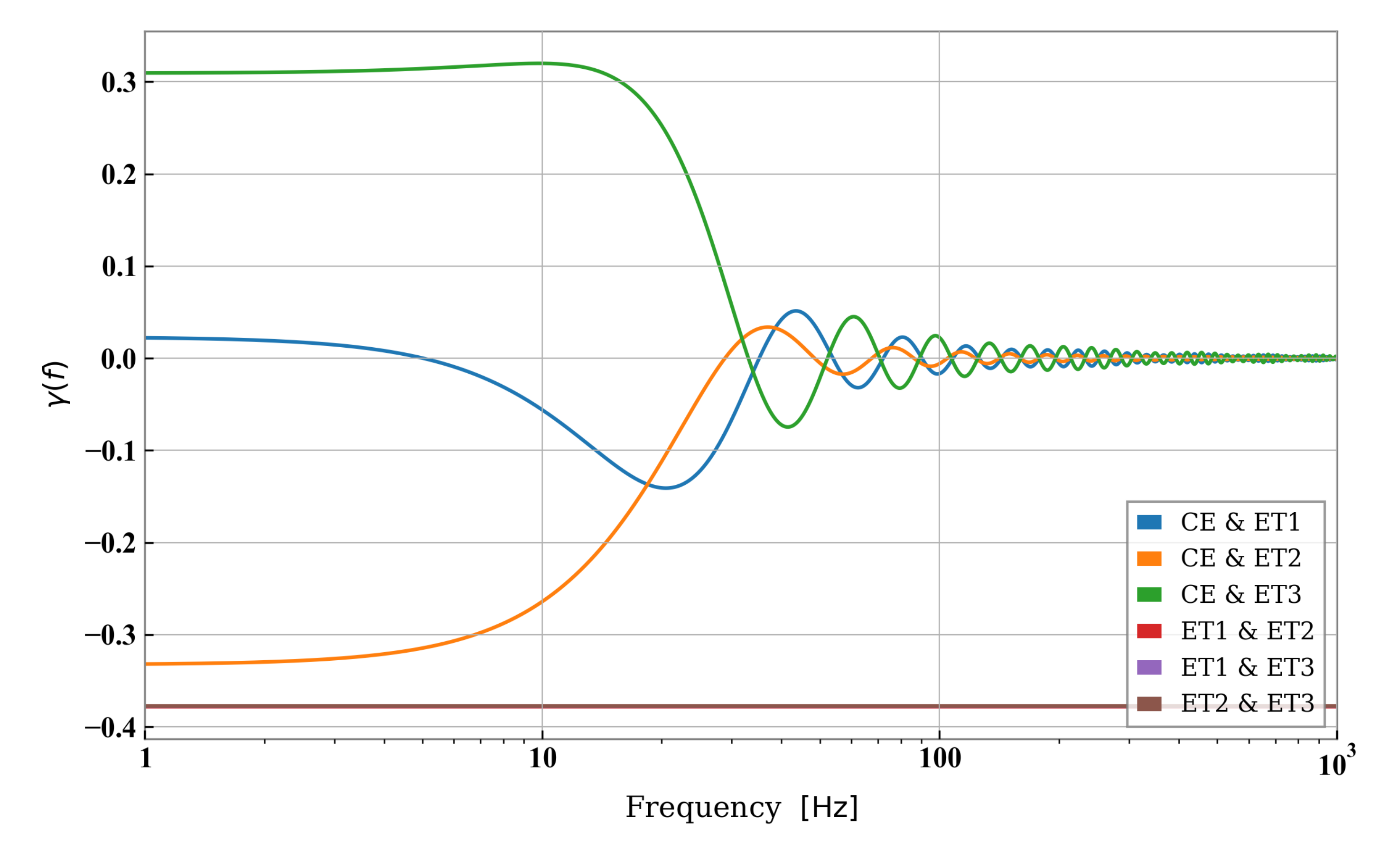}
	\caption{ORF $\gamma(f)$ between Cosmic Explorer (CE) and Einstein Telescope (ET). The ORFs are shown over a logarithmic (top) and linear (bottom) frequency axis. Note that the ORF between different detectors of the ET triangle is constant with a value of about -0.38.} 
	\label{fig:ORF}
\end{figure}

The ORFs between CE and ET are shown in figure \ref{fig:ORF}. While correlation measurements between detectors of the ET triangle are sensitive to stochastic backgrounds over ET's entire observation band, correlation measurements between CE and ET are most sensitive only up to about 20\,Hz. However, correlating between ET detectors bears a much greater risk that other than GW signals, e.g., local magnetic and seismic disturbances, cause additional correlated contributions, which might limit ET's sensitivity as stand-alone observatory of stochastic GW backgrounds. The ET-only sensitivity will greatly depend on cancellation techniques for environmental noise as proposed in \cite{Cel2000,CoEA2018c}, or the inclusion of ET's GW null-stream \cite{Tania2012}.

With the definition of the ORF in equation (\ref{eq:ORF}), the CPSD between two detectors due to the stochastic GW background can be written \cite{Chiara2019}
\begin{equation}
C_{ij}(f)  = S_{\rm GW}(f) \gamma_{ij}(f), \quad S_{\rm GW}(f) = \frac{3H_{0}^{2}}{10 \pi^{2}} \cdot \frac{\Omega_{\rm GW}(f)}{f^{3}}.
\label{eq:cpsdGW}
\end{equation}
This value needs to be confronted with the average statistical error of the CPSD from uncorrelated instrument noise,
\begin{equation}
\sigma_{ij}(f) = \sqrt{\frac{S_{i}(f) S_{j}(f)}{N}},
\label{eq:sigma}
\end{equation}
where $S_i(f)$ is the instrument noise spectral density, and $N$ is the total number of averages going into the estimate of the CPSD. For example, if the total time-stretch of data is $T$, and the CPSD is calculated using segments of length $\tau$ for the fast Fourier transforms (FFTs), and the CPSD calculation foresees the application of spectral windows (anti-leakage), which means that something like 50$\%$ overlap between FFT segments is recommended to make full use of all the information in the data, then we have $N \approx 2T/\tau$. 

\subsection{Optimal filter}
\label{subsec:OpF}
The optimal search for a stochastic background with known or modeled spectral shape involves the integral of CPSDs over frequency. However, since the relative contributions of the stochastic signal and instrument noise to the CPSD vary over frequency, the optimal integration should use a filter $\tilde{Q}_{ij}(f)$, which emphasizes some parts of the spectrum over others. 

\begin{figure}
	\centering
	\includegraphics[width=\columnwidth]{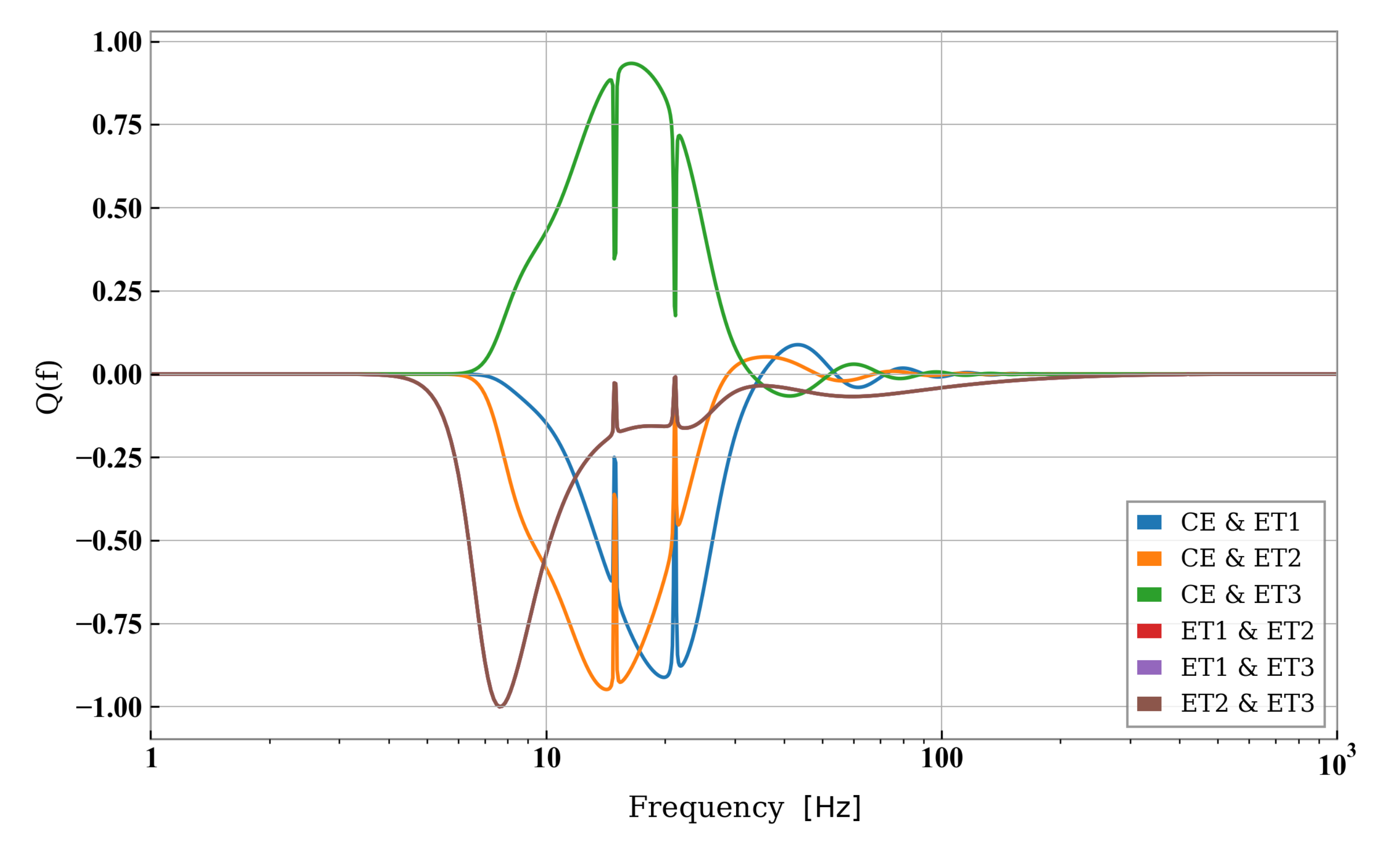}
	\caption{Optimal filter function $\tilde{Q}(f)$ between CE and ET plotted over a logarithmic (top) and linear (bottom) frequency axis. The optimal filters between detectors of the ET triangle are all identical.}
	\label{fig:optimalfilt}
\end{figure}

The signal-to-noise ratio (SNR) of a filtered search is determined by the mean value of the integrated CPSD signals \cite{Bruce1999}
\beq
\begin{split}
\langle \mathcal C_{ij}\rangle &= \int\limits_0^\infty\drm f\, \langle C_{ij}(f)\rangle \tilde{Q}_{ij}(f)\\
&= \int\limits_0^\infty\drm f\, \gamma_{ij}(f)S_{\rm GW}(f) \tilde{Q}_{ij}(f),
\end{split}
\label{eq:meancsd}
\eeq
and their variances
\beq
\langle (\mathcal C_{ij})^2\rangle = \frac{1}{2T}\int\limits_0^\infty\drm f\, S_i(f)S_j(f)|\tilde{Q}_{ij}(f)|^2.
\label{eq:varcsd}
\eeq
The averages are over many independent estimates of CPSDs. It is straight-forward to show that the optimal filter function is given by
\begin{equation}
\tilde{Q}_{ij}(f) = \mathcal{N} \frac{\gamma_{ij}^*(f)S_{GW}(f)}{S_i(f) S_j(f)},
\label{eq:optimalfilt}
\end{equation}
where $\mathcal{N}$ is a normalization factor, which has no influence on the SNR. The form of the optimal filters (in arbitrary, but consistent normalization) is shown in figure \ref{fig:optimalfilt}.
In all cases, the optimal filter emphasizes contributions from low frequencies near the lower bound of the observation band of the GW detectors.

Inserting the filter into the previous two equations, we obtain
\beq
{\rm SNR}^2=\frac{\langle\mathcal C_{ij}\rangle^2}{\langle (\mathcal C_{ij})^2\rangle}=2T\int\limits_0^\infty\drm f\,\frac{|\gamma_{ij}(f)|^2S_{\rm GW}^2(f)}{S_1(f)S_2(f)}
\label{eq:stochSNR}
\eeq
Note that in a discrete version of this equation, the integral becomes a sum over all positive frequency bins, and the $\drm f$ needs to be replaced by $1/\tau$. 
\begin{figure}
	\centering
	\includegraphics[width=\columnwidth]{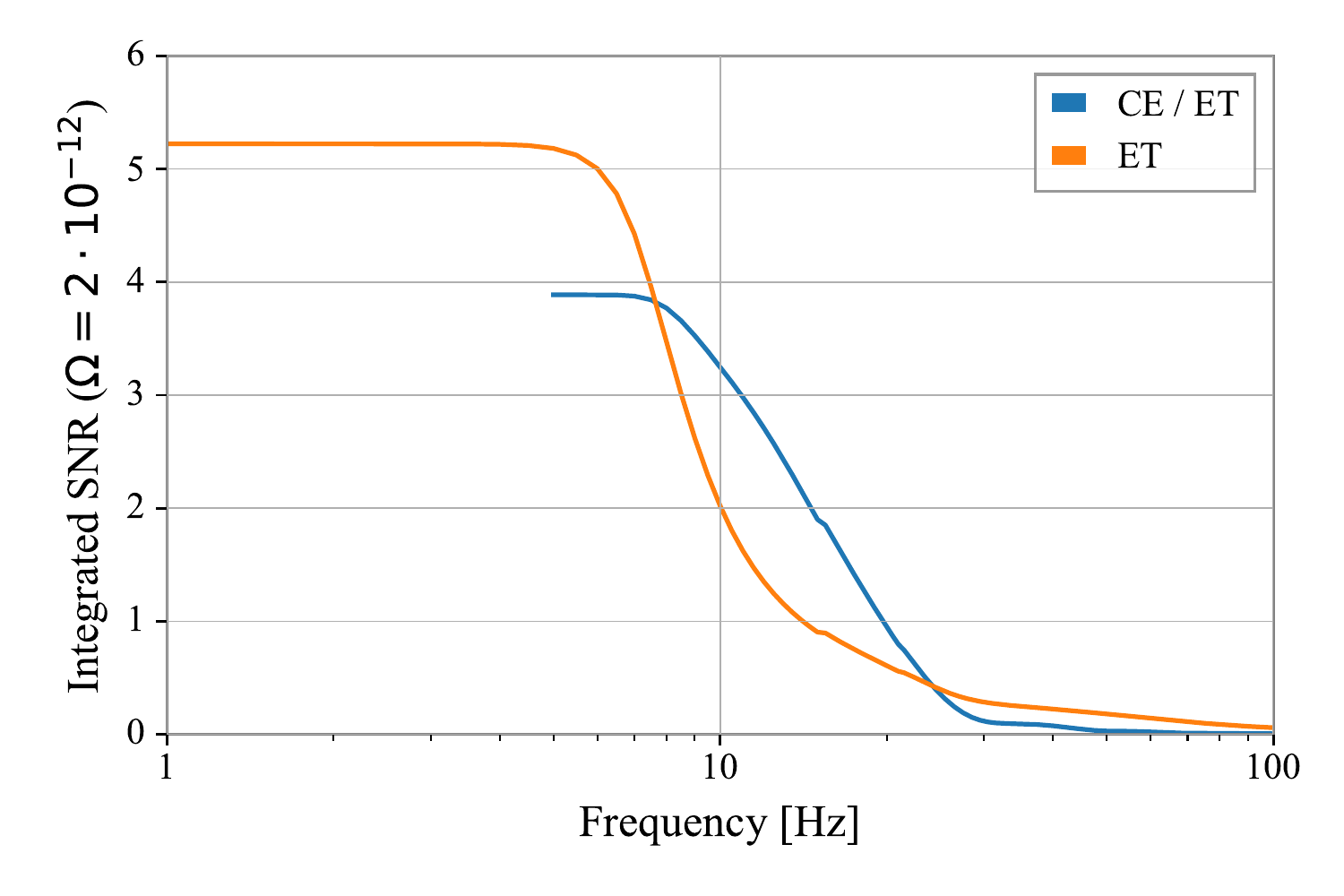}
	\caption{Signal-to-noise ratio of a flat $\Omega_{\rm GW}=2\cdot10^{-12}$ stochastic background. The curves are the SNRs accumulated from high to low frequencies. Total observation time is 1.3 years.}
	\label{fig:specSNR}
\end{figure}
Figure \ref{fig:specSNR} shows the SNR of a flat-$\Omega=2\cdot10^{-12}$ stochastic background observed over 1.3 years with CE and ET. The curves represent the SNRs accumulated from high to low frequencies, such that the lowest frequency values shown in the plot correspond to the SNR of the correlation measurements making use of all three detectors of an ET triangle. In this way, it is possible to see, at which frequencies most of the SNR is accumulated. Cosmic Explorer correlated with ET is most sensitive to a flat background between 8\,Hz and 30\,Hz, while ET by itself accumulates its SNR over a slightly broader band. The total SNR achieved by ET in this case is 5.2, while CE correlated with ET achieves an SNR of 3.9. 

\section{Projection results}
\label{sec:results}
The goal is to demonstrate that subtraction residuals can limit the sensitivity of 3G detectors to a CGWB and that the noise-projection method can remove subtraction residuals. In other words, we need to show that subtraction residuals can lie above the instrument-noise contribution of equation (\ref{eq:sigma}), and that projection suppresses residuals to a level significantly below the instrument noise. 

\begin{figure*}[!htbp]
	\centering
	\includegraphics[width=\columnwidth]{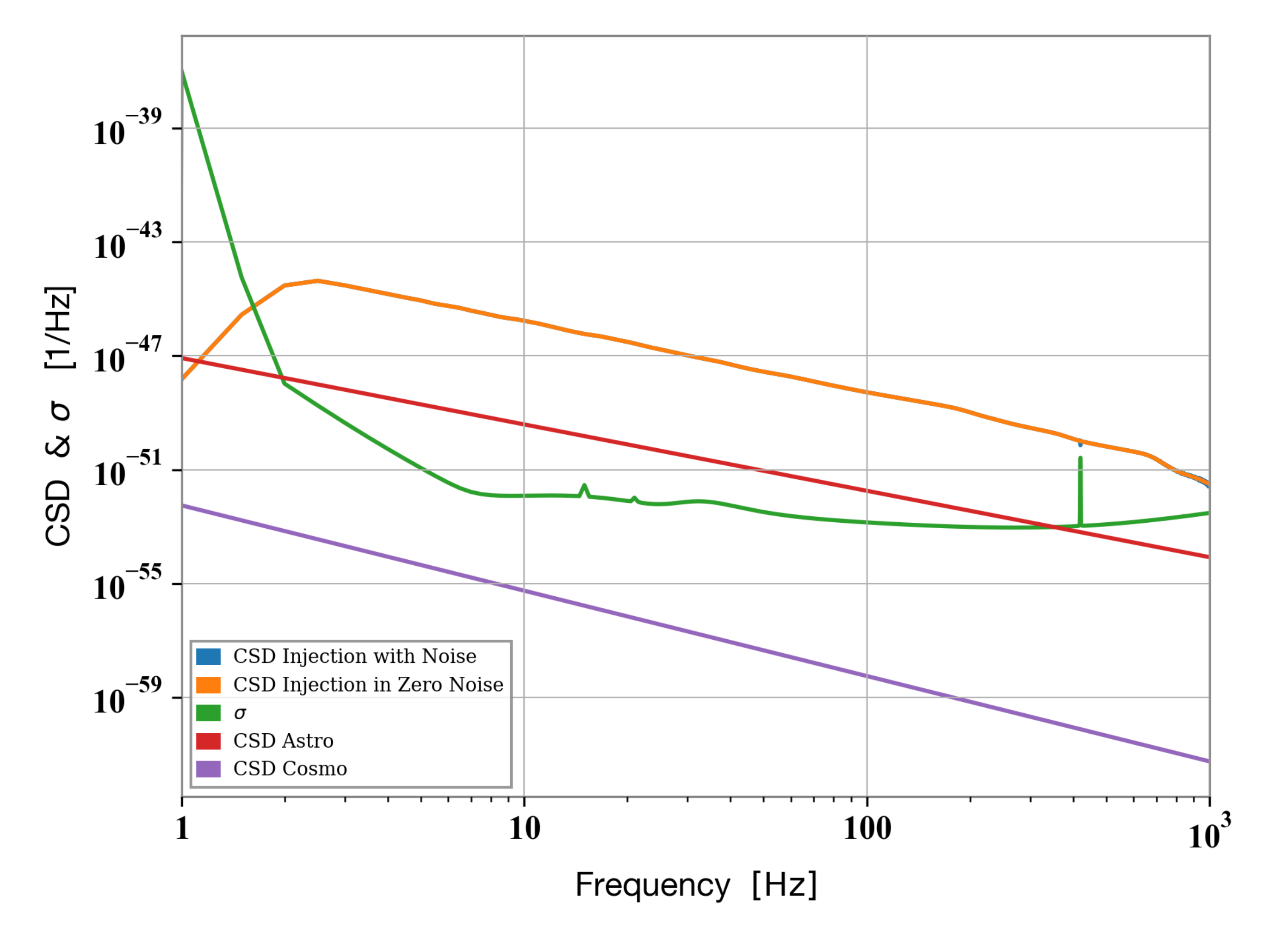}
	\includegraphics[width=\columnwidth]{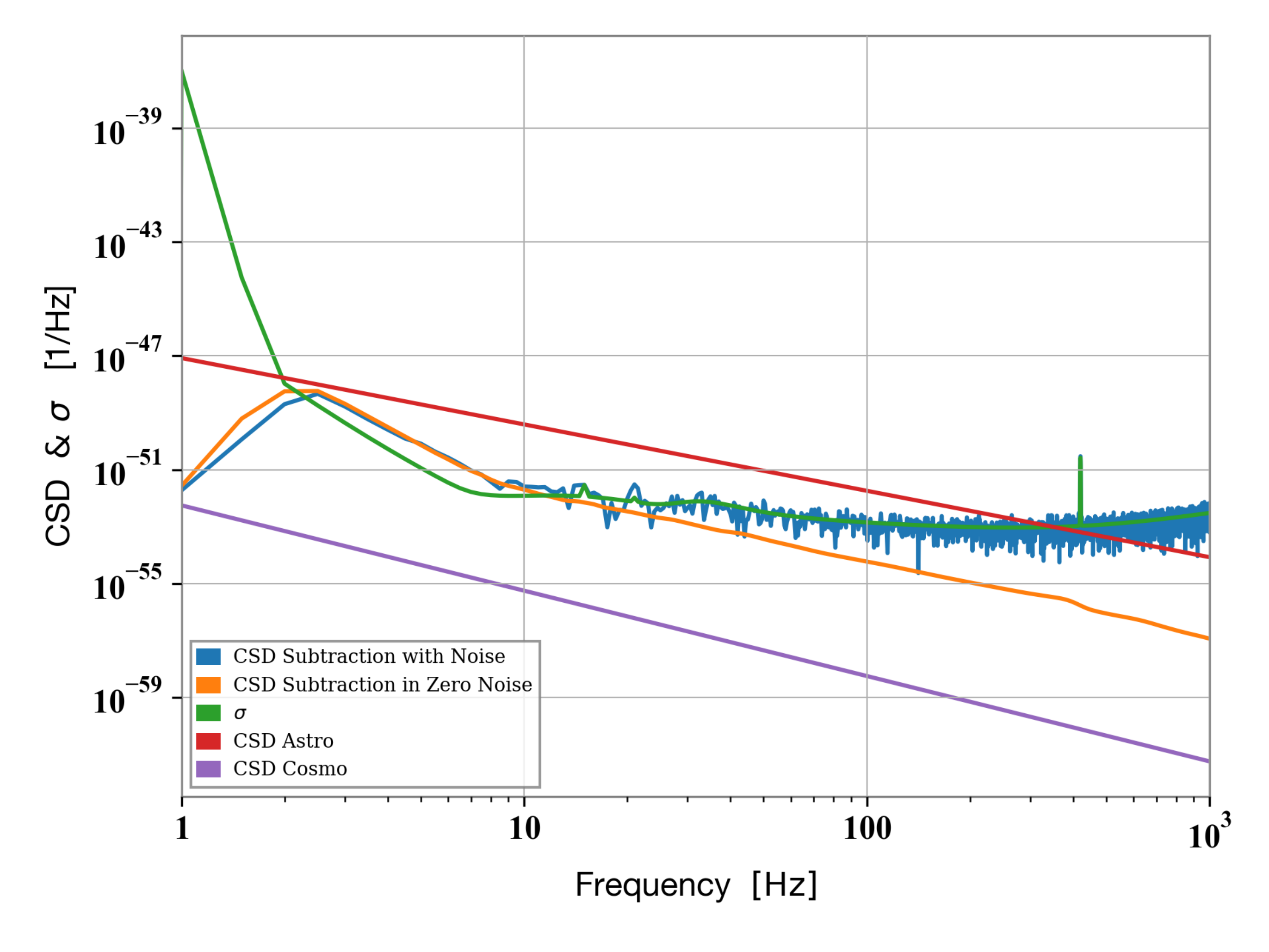}
	\includegraphics[width=\columnwidth]{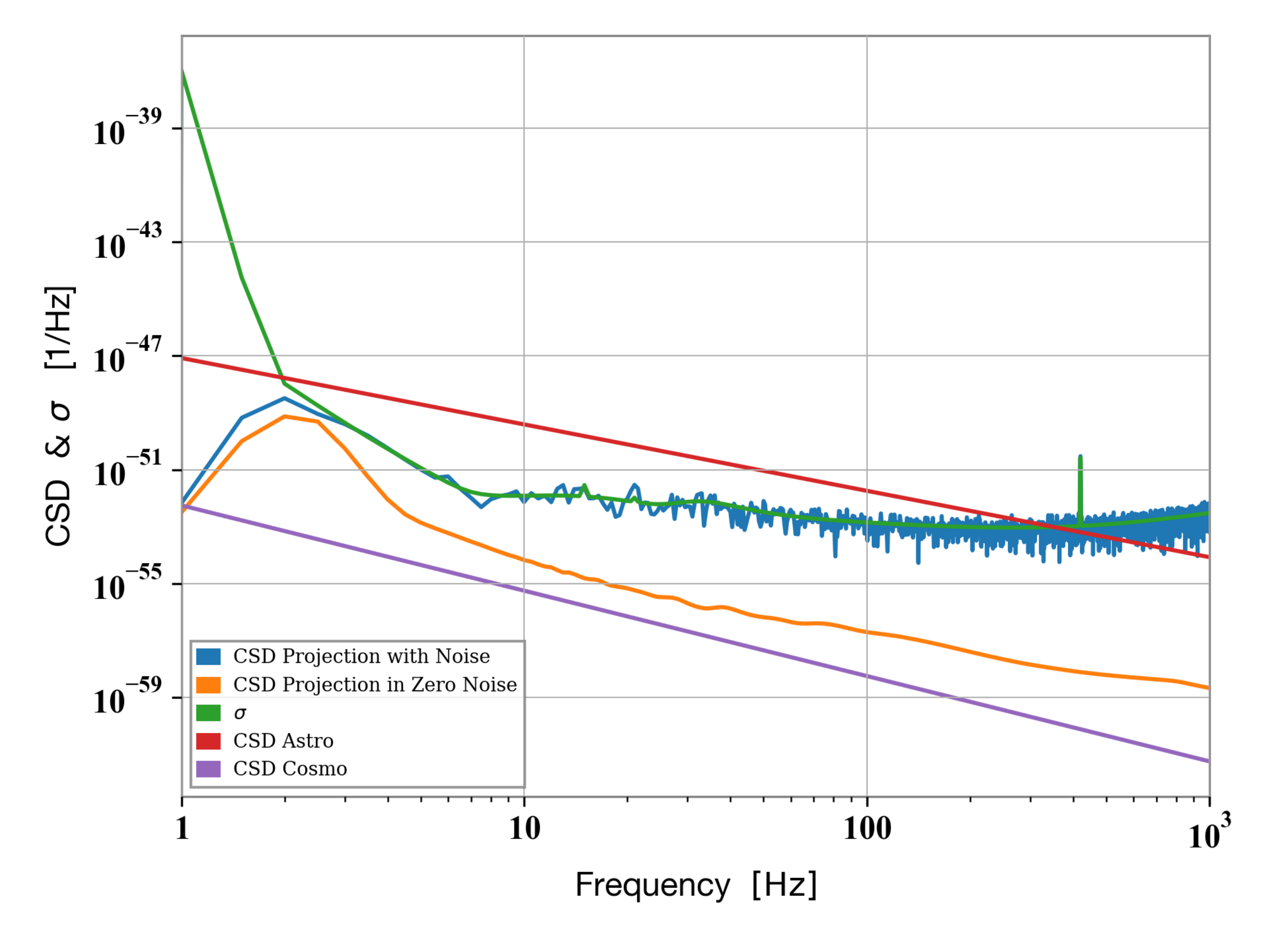}
	\caption{Plots of residual CSDs averaged over all ET detector pairs with BBH foreground (top, left), after best-fit subtraction (top, right), and after noise projection (bottom). The CPSDs are 1.3-year averages and also averaged over all detector pairs of the ET triangle with a total of $10^6$ injected BBHs. The astrophysical reference model (red curves) is only an approximation valid below about 100\,Hz since it is predicted to fall more strongly above 100\,Hz. The purple curve represents a CGWB with frequency independent $\Omega_{\rm GW}=10^{-15}$. The blue curves are the simulated CPSD measurements. The green curves are the predicted instrument noise. The orange curves show the CPSDs for simulations without instrument noise.}
	\label{fig:residualCSDs}
\end{figure*}

We focus this analysis on ET. The CPSDs are calculated from $\tau=2\,$s discrete Fourier transforms using the Welch method with 50\% overlap between segments. As stated before, the total simulated time is 1.3 years or $T=4\times 10^7$\,s. The CPSDs are averaged over all three ET detector pairs. The results are shown in figure \ref{fig:residualCSDs}.

The plots contain reference models of the astrophysical foreground with $\Omega_{GW} =7 \times 10^{-10} \times (f/10 \rm{Hz})^{2/3}$, which approximates past estimates \cite{Tania2017}, and a CGWB with frequency independent $\Omega_{GW} = 10^{-15}$. 

%\rednote{The cosmological background is fine. However, there is still the question about the astrophysical foreground. In Tania's paper, they used 7e-10 at 10Hz, while you have 5e-9 in your code. So, if you can find a reference for the 5e-9, then we can cite it and keep plots in figure 8 as they are. If not, then we leave the citation to Tania's paper, but then the red curves in figure 8 need to be recalculated. Just to make sure, note that 1e-10 means $10^{-10}$.}

The upper, left plot shows the CPSDs before subtraction of the foreground, the upper, right plot after subtraction, and the bottom plot after projection. The instrument noise of the CPSD (green curve) is calculated using equation (\ref{eq:sigma}). In all three plots, the orange curves are the CPSDs from simulations without instrument noise. 

The astrophysical BBH foreground shown in the top, left plot (blue curve, hidden behind orange curve) exceeds past predictions (red curve). This is mostly due to the fact that we selected higher-SNR members of the BBH population, for which we expect a Fisher-matrix based projection method to work efficiently. The subtraction residuals in the top, right plot lie above the instrument noise below 10\,Hz. It confirms that the sensitivity of ET to a CGWB can be limited by subtraction residuals. This is true for 1.3 years of observation time, and remains true for longer observation times (increasing observation time lowers the instrument noise in these plots, and leaves all other curves the same). Since the spectrum of subtraction residuals depends weakly on the SNRs of the members of the astrophysical foreground (as long as the BBHs can be detected), this conclusion remains valid for more realistic models of the astrophysical foreground. The impact of low-SNR signals, of which only some are detected, or which are included as sub-threshold signal candidates in the subtraction, projection procedure needs to be investigated in future work. The projected residuals (blue curve) in the bottom plot are fully consistent with the instrument-noise model, which means that subtraction residuals were successfully reduced. The full potential of a CGWB search with ET is restored, at least with respect to the higher-SNR signals of a BBH population.

\section{Conclusion}
In this paper, we presented an analysis of a noise-projection method based on a higher-order geometrical analysis of matched-filter GW searches to mitigate subtraction residuals of an astrophysical foreground in the proposed third-generation detectors Einstein Telescope and Cosmic Explorer. We showed that the projection method can improve the sensitivity to a CGWB. We provided insight into why the projection method is expected to work, and we tested the method with a time-domain simulation of a future detector network. The important first step of the analyses, i.e., the estimation of BBH parameters, was carried out with a state-of-the-art parameter-estimation software (Bilby) by posterior sampling. The presented results are a proof-of-principle since some simplifications of the simulation of the astrophysical foreground had to be done.

The results indicate that the projection method is able to remove all influence of subtraction residuals from BBHs on searches of a CGWB. However, two important aspects need to be addressed in future work. First, the impact of low-SNR signals in the astrophysical foreground on the sensitivity of CGWB searches needs to be investigated. Some of these signals will be visible as sub-threshold signals, others complete hidden in instrumental noise. Their contribution to the astrophysical foreground must be sufficiently low to not pose a fundamental limit to the capacity ET and CE have for CGWB observations. Second, since the foreground removal requires signal models, the dependence of the residuals on choices of waveform models needs to be assessed. Since our implementation of the projection method is fully numerical, we do not require analytical expressions for the waveform models to calculate the projection operators. 

\section{Acknowledgement}
We are grateful to Cristiano Palomba for providing helpful comments on the manuscript. We acknowledge the use of  inference library Bilby for parameter estimation and detector modelling. 

\bibliographystyle{apsrev4-2} 
\bibliography{references}
\end{document}